\documentclass[prd,preprint,letterpaper]{revtex4}
\usepackage{graphicx}
\usepackage{amsmath,amssymb}
\input{epsf}
\usepackage{epsf}
\newcommand {\ga} {\ {\raise-.5ex\hbox{$\buildrel>\over\sim$}}\ }
\newcommand {\la} {\ {\raise-.5ex\hbox{$\buildrel<\over\sim$}}\ }

\begin{document}

\def\be{\begin{equation}}
\def\ee{\end{equation}}

\title{The Return of a Static Universe and the End of Cosmology}
\author{Lawrence M. Krauss$^{1,2}$ and Robert J. Scherrer$^{2}$}
\affiliation{$^1$Department of Physics, Case Western Reserve University,
Cleveland, OH~~44106;  email: krauss@cwru.edu}
\affiliation{$^2$Department of Physics \& Astronomy, Vanderbilt University,
Nashville, TN~~37235; email: robert.scherrer@vanderbilt.edu}
\date{\today}

\begin{abstract}
We demonstrate that as we extrapolate the current $\Lambda$CDM universe forward in time, all evidence of the Hubble expansion will
disappear, so that observers in our ``island universe" will be fundamentally incapable of determining the true nature
of the universe, including the existence of the highly dominant vacuum energy, the existence of the CMB, and
the primordial origin of light elements.  With these pillars of the modern Big
Bang gone, this epoch will mark 
the end of cosmology and the return of a static universe.  In this sense, the
coordinate system appropriate for future observers will perhaps fittingly
resemble the static coordinate system in which the de Sitter universe was first presented.
\end{abstract}

\maketitle

Shortly after Einstein's development of general relativity, the Dutch astronomer
Willem de Sitter proposed  a static model of the universe containing no matter, which he thought might be a reasonable approximation to our low density universe.   
One can define a coordinate system in which the de Sitter metric takes a static form by defining de Sitter spacetime with a cosmological constant $\Lambda$ as a four dimensional hyperboloid $ {\cal S}_\Lambda: \eta^{}_{AB} \xi^A \xi^B = -R^2,\qquad R^2=3\Lambda^{-1} $ embedded in a $5d$ Minkowski spacetime with 
  	  	$ds^2=\eta^{}_{AB} d\xi^A d\xi^B, \quad$ 	and
  	  	$ (\eta_{AB})={\rm diag}(1, -1, -1, -1, -1), \qquad {A, B=0,\cdots,4}.$ 	
The static form of the de Sitter metric is then
$$ ds_s^2 = (1-r_s^2/R^2)dt_s^2 - \frac {dr_s^2}{1-r_s^2/R^2}-r_s^2d\Omega^2,$$
 which can be obtained by setting
                          $\displaystyle \xi^0$ 	= 	$\displaystyle (R^2-r_s^2)^{1/2}\sinh (t_s/R),$ 	 
$\displaystyle \xi^1$ 	= 	$\displaystyle r_s \sin \theta \cos \varphi,$ 	 
$\displaystyle \xi^2$ 	= 	$\displaystyle r_s \sin \theta \sin \varphi,$
$\displaystyle \xi^3$ 	= 	$\displaystyle r_s \cos \theta,$ 	 
$\displaystyle \xi^4$ 	= 	$\displaystyle (R^2-r_s^2)^{1/2}\cosh (t_s/R).$ 
 In this case the metric only corresponds to the section of de Sitter space within a cosmological horizon at $R=r-s$.

In fact de Sitter's model wasn't globally static, but eternally expanding, as can be seen by a coordinate transformation which explicitly incorporates the time dependence of the scale factor $R(t) = \exp(Ht)$.  While spatially flat, it actually incorporated Einstein's cosmological term, which is of course now understood to be equivalent to a vacuum energy density,  leading to a redshift proportional to distance.   

The de Sitter model languished for much of the last century, once the Hubble expansion had been discovered, and the cosmological term abandoned. However, all present observational evidence is consistent with a $\Lambda$CDM flat universe
consisting of roughly 30\% matter (both dark matter and baryonic matter)
and 70\% dark energy  \cite{kraussturn,perlm,reiss}, with the latter having a density that appears constant
with time.  All cosmological models with a non-zero cosmological constant will approach a de Sitter universe in the far future, and many of the
implications of this fact have been explored in the literature \cite{KS,Star,GB,Loeb1,Chiueh,Busha,NL1,NL2,Heyl,decay}.

Here we re-examine the practical significance of the ultimate de Sitter expansion and point out a new
eschatological physical consequence:  from the perspective of any observer within a bound gravitational system in the far future, the static version of de Sitter space outside of that system will eventually become the appropriate physical coordinate system.  Put more succinctly, in a time comparable to the age of the longest lived stars,  observers will not be able to perform any observation or experiment that infers either the existence of an
expanding universe dominated by a cosmological constant, or that there was a hot Big Bang.  Observers will be able to infer a finite
age for their island universe, but beyond that cosmology will effectively be over.  The static universe, with which cosmology at the
turn of the last century began, will have returned with a vengeance.

Modern cosmology is built on integrating general relativity and three observational pillars:
the observed Hubble expansion, detection of the cosmic
microwave background radiation, and the determination of the abundance of elements
produced in the early universe.  We describe next in detail how these observables will disappear for an observer in the far future, and how this will be likely to affect the theoretical conclusions one might derive about the universe.

\subsection{The disappearance of the Hubble Expansion}

The most basic component of modern cosmology is the expansion
of the universe, firmly established by Hubble in 1929.  Currently, galaxies and
galaxy clusters are gravitationally bound and have dropped out of the
Hubble flow, but structures on larger length scales are observed to obey
the Hubble expansion law.  Now consider what happens in the far future
of the universe.  Both analytic \cite{Loeb1} and numerical
\cite{NL1} calculations indicate that the Local Group remains gravitationally
bound in the face of the accelerated Hubble expansion.  All
more distant
structures will be driven outside of the de Sitter event horizon in a timescale
on the order of 100 billion years (\cite{KS}, see also Refs. \cite{Chiueh,Busha}).  While objects will not be observed to cross the event horizon, light from them will be exponentially redshifted, so that within a time frame comparable to the longest lived main sequence stars all objects outside of our local cluster will truly become invisible \cite{KS}. 

Since the only remaining visible objects will in fact be gravitationally bound and decoupled from the underlying Hubble
expansion, any local observer in the far future will see a single galaxy
(the merger product of the Milky Way and Andromeda and other remnants of the
Local Group) and will have
no observational evidence of the Hubble expansion.  Lacking such
evidence, one may wonder whether such an observer will postulate
the correct cosmological model.  We would argue that in fact, such an observer will conclude the existence of a static ``island universe,"
precisely the standard model of the universe c. 1900.

This will be true in spite of the fact that the dominant energy in this universe will
not be due to matter, but due to dark energy, with $\rho_M/\rho_\Lambda \sim 10^{-12}$ inside
the horizon volume \cite{Busha}.
The irony, of course, is that the denizens of this static universe
will have no idea of the existence of the dark energy, much
less of its magnitude, since they will have no probes of the length
scales over which $\Lambda$ dominates gravitational dynamics.  It appears that
dark energy is undetectable not only in the limit where $\rho_\Lambda \ll \rho_M$, but
also when $\rho_\Lambda \gg \rho_M$.

Even if there were no direct evidence of the Hubble expansion, we might expect
three other bits of evidence, two observational and one theoretical, to lead physicists in the future to ascertain the underlying nature of cosmology.  However, we next describe how this is unlikely to be the case.

\subsection{Vanishing CMB}

The existence of a Cosmic Microwave Background was the key observation that convinced most physicists and astronomers that there was in fact a hot big bang, which essentially implies a Hubble expansion today.   But even if skeptical observers in the future
were inclined to undertake a search for this afterglow of the Big Bang, they would
come up empty-handed.  At $t \approx$ 100 Gyr, the peak
wavelength of the cosmic microwave background will be redshifted
to roughly $\lambda \approx 1$ m, or a frequency of roughly
300 MHz.  While a uniform radio background at this frequency would
in principle be observable, the intensity of the CMB will also
be redshifted by about 12 orders of magnitude.  At much later times, the CMB becomes
unobservable even in principle, as the peak wavelength is driven
to a length larger than the horizon \cite{KS}.  Well before then, however,
the microwave background peak will redshift below
the plasma frequency of the interstellar medium, and so will be screened from any observer within the galaxy.
Recall that the plasma frequency is given by
$$\nu_p = \left(\frac{n_e e^2}{\pi m_e}\right)^{1/2},$$
where $n_e$ and $m_e$ are the electron number density and mass, respectively.  Observations of dispersion in pulsar
signals give \cite{ism} $n_e \approx 0.03$ cm$^{-3}$ in the interstellar medium, which corresponds to a plasma
frequency of $\nu_p \approx 1$ kHz, or a wavelength of $\lambda_p \approx 3 \times 10^7$ cm.  This corresponds
to an expansion factor $\sim 10^8$ relative to the present-day peak of the CMB.  Assuming an exponential expansion, dominated by dark energy, this expansion factor will be reached when the universe is less than 50 times its present age, well below the lifetime of the longest-lived main sequence stars.

After this time, even if future residents of our island universe set out to measure a universal
radiation background, they would be unable to do so.  The wealth of information about early universe cosmology that can be derived from fluctuations in the CMB would be even further out of reach.

\subsection{General Relativity Gives No Assistance}

We may assume that theoretical physicists in the future will infer that
gravitation is described by general relativity, using observations of planetary
dynamics, and ground-based tests of such phenomena as gravitational time
dilation.  Will they then not be led to a Big Bang expansion, and a beginning
in a Big Bang singularity, independent of data, as Lemaitre was?   Indeed, is
not a static universe incompatible with general relativity?  

The answer is no.  The inference that the universe must be expanding or
contracting is dependent upon the cosmological hypothesis that we live in an
isotropic and homogeneous universe.  For future observers, this will manifestly
not be the case.   Outside of our local cluster, the universe will appear to be
empty and static.   Nothing is inconsistent with the temporary existence of a
non-singular isolated self-gravitating object in such a universe, governed by
general relativity.   Physicists will infer that this system must ultimately
collapse into a future singularity, but only as we presently conclude our galaxy must ultimately coalesce into a large black hole. Outside of this region, an empty static universe can prevail.

While physicists in the island universe will therefore conclude that their island has a finite future, the question will naturally arise as to whether it had a finite beginning.  As we next describe, observers will in fact be able to determine the age of their local cluster, but not the nature of the beginning.

\subsection{Polluted Elemental Abundances}

The theory of Big Bang Nucleosynthesis reached a fully-developed state
\cite{Wagoner}
only after the discovery of the CMB (despite early abortive attempts
by Gamow and his collaborators \cite{Gamow}).  Thus, it is unlikely
that the residents of the static universe would have
any motivation to explore the possibility of primordial nucleosynthesis.
However, even if they did,
the evidence for BBN rests crucially
on the fact that relic abundances of deuterium remain observable at
the present day, while helium-4 has been enhanced by only a few percent
since it was produced in the early universe.  Extrapolating forward
by 100 Gyr, we expect significantly more contamination of the helium-4
abundance, and concomitant destruction of the relic deuterium.  It has been
argued \cite{Adams} that the ultimate extrapolation of light elemental abundances,
following many generations of stellar evolution, is a mass fraction of helium given by
$Y=0.6$.  The primordial
helium mass fraction of $Y=0.25$ will be a relatively small fraction of this
abundance.  It is unlikely that much deuterium could survive this degree of processing.
Of course, the current ``smoking gun" deuterium abundance is provided by Lyman-$\alpha$
absorption systems, back-lit by QSOs (see, e.g., Ref. \cite{Kirkman}).
Such systems will be unavailable to our observers of the future, as both the QSOs and the
Lyman-$\alpha$ systems will have redshifted outside of the horizon.

Astute observers will be able to determine a lower limit on the age of their system, however, using standard stellar
evolution analyses of their own local stars.   They will be able to examine the locus of all stars and extrapolate
to the oldest such stars to estimate a lower bound on the age of the galaxy.   They will be able to determine an
upper limit as well, by determining how long it would take for all of the
observed helium to be generated by stellar nucleosynthesis.   However, without
any way to detect primordial elemental abundances, such as the aforementioned possibility of measuring deuterium in distant intergalactic clouds that currently absorb radiation from distant quasars and allow a determination of the deuterium abundance in these pre-stellar systems, and with the primordial helium abundance dwarfed by that produced in stars, inferring the original BBN abundances will be difficult, and probably not well motivated. 

Thus, while physicists of the future will be able to infer that their island universe has not been eternal, it is
unlikely that they will be able to infer that the beginning involved a Big Bang.

\subsection{Conclusion}

The remarkable cosmic coincidence that we happen to live at the only time in the history of the universe when the
magnitude of dark energy and dark matter densities are comparable has been a source of great current speculation,
leading to a resurgence of interest in possible anthropic arguments limiting the value of the vacuum energy (see,
e.g., Ref. \cite{Weinberg}).  But this coincidence endows our current epoch with another special feature, namely that we can actually infer both the existence of the cosmological expansion, and the existence of dark energy.  Thus, we live in a very special time in the evolution of the universe: the time at which we can observationally verify that we live in a very special time in the evolution of the universe!  

Observers when the universe was an order of magnitude younger would not have been able to discern any effects of dark energy on the expansion, and observers when the universe is more than an order of magnitude older will be hard pressed to know that they live in an expanding universe at all, or that the expansion is dominated by dark energy.  By the time the longest lived main sequence stars are nearing the end of their lives, for all intents and purposes, the universe will appear static, and all evidence that now forms the basis of our current understanding of cosmology will have disappeared. 

\vskip 0.2in
{\it Note added in proof: } After this paper was submitted we learned of a prescient 1987 paper \cite{Ellis},
written before the discovery of dark energy and other cosmological observables  that are central to our analysis, which nevertheless  raised the general question of whether there would be epochs in the Universe when observational
cosmology, as we now understand it, would not be possible.

\acknowledgments 
L.M.K. and R.J.S. were supported in part by the Department of Energy.

\end{document}